\documentclass[preprint,12pt]{elsarticle}

\usepackage{amsmath,euscript,amssymb,epsfig,amsfonts,latexsym}
\usepackage{graphicx,color}

\newcommand{\be}[1]{\begin{equation}\label{#1}}
\newcommand{\ee}{\end{equation}}
\newcommand{\ba}[1]{\begin{eqnarray}\label{#1}}
\newcommand{\ea}{\end{eqnarray}}
\newcommand{\rf}[1]{(\ref{#1})}
\newcommand{\nn}{\nonumber}

\journal{Physics of the Dark Universe}

\begin{document}

\begin{frontmatter}

\title{Perfect fluids with $\omega=\mathrm{const}$ as sources\\ of scalar cosmological perturbations}

\author{Maxim Eingorn}\ead{maxim.eingorn@gmail.com}

\address{North Carolina Central University, CREST and NASA Research Centers\\ Fayetteville st. 1801, Durham, North Carolina 27707, U.S.A.}

\author{Ruslan Brilenkov}\ead{ruslan.brilenkov@gmail.com}

\address{Institute for Astro- and Particle Physics, University of Innsbruck\\ Technikerstrasse 25/8, A-6020 Innsbruck, Austria}

\address{Dipartimento di Fisica e Astronomia `G. Galilei', Universit\`{a} di Padova\\ vicolo dell'Osservatorio 3, 35122 Padova, Italy}

\begin{abstract}
We make a generalization of a self-consistent first-order perturbation scheme, being suitable for all (sub-horizon and super-horizon) scales, which has been
recently constructed for the concordance cosmological model and discrete presentation of matter sources, to the case of extended models with extra perfect fluids
and continuous presentation. Namely, we derive a single equation determining the scalar perturbation and covering the whole space as well as define the
corresponding universal Yukawa interaction range. We also demonstrate explicitly that the structure growth is suppressed at distances exceeding this fundamental
range.
\end{abstract}


\begin{keyword}
inhomogeneous Universe \sep scalar cosmological perturbations \sep gravitational potential \sep Yukawa interaction
\end{keyword}

\end{frontmatter}

\

\section{Introduction}

The conventional cosmological model satisfies the modern observational data and describes the Universe filled with prevailing dark energy (represented by the
cosmological constant) and cold dark matter (CDM) as well as standard baryonic matter and radiation (playing secondary roles). Along with this paradigm, there are
a number of alternatives, which assume the presence of some additional Universe constituent, e.g., in the form of a perfect fluid with constant parameter $\omega$
in the linear equation of state (EoS). The cosmological constant itself may be interpreted as such a fluid with $\omega=-1$ (or, in other words, with vacuum-like
EoS). If $\omega\neq-1$, but the parameter $\omega$ is close enough to $-1$, then this fluid may be used instead of the cosmological term in order to ensure the
late-time acceleration (see, particularly, \cite{Planck13,Planck15} for the corresponding experimental constraints imposed on the dark energy EoS of the specified
form). At the same time, there can be components with $\omega$ being quite far from $-1$. For example, frustrated networks of topological defects (cosmic strings
and domain walls) have the form of perfect fluids with constant parameters $\omega=-1/3$ and $\omega=-2/3$, respectively \cite{ShellVil,Avelino,Kumar,Cappi}. In
general, depending on the comparison of the negative parameter $\omega$ and the vacuum value $-1$, these components are usually called quintessence
\cite{quintess1,quintess2} and phantom \cite{phantom1,phantom2,phantom3} for $-1<\omega<0$ and $\omega<-1$, respectively.

In compliance with the cosmological principle, the Universe is considered homogeneous and isotropic and described by the corresponding background
Friedmann-Lema\^{\i}tre-Robertson-Walker (FLRW) metric at sufficiently large scales. Meanwhile, at small enough scales the Universe is apparently highly
inhomogeneous (galaxies, groups and clusters of galaxies are observed). In the recent paper \cite{Eingorn} the unified first-order perturbation scheme, being
valid for arbitrary (sub-horizon and super-horizon) distances, 
was developed in the framework of the $\Lambda$CDM model in the weak gravitational field limit. This scheme has a number of successes in resolving different
challenges and promises to be important for high-precision cosmology and $N$-body simulations covering huge spatial volumes. Consequently, it makes sense to
generalize this approach, elaborated for the presentation of nonrelativistic matter in the form of discrete gravitating particles (see also the preceding papers
\cite{EZcosm1,EKZ2,EZcosm2}), to the case of the continuous presentation of sources, namely, the standard $\Lambda$CDM components in combination with
supplementary inhomogeneous fluids undergoing adiabatic perturbations. We make this necessary generalization in the current paper and thereby provide an ample
opportunity to investigate the structure formation and growth within the nonconventional cosmological models at arbitrary scales. This can help to distinguish
among them and the concordance paradigm.

The paper is organized in the following way. In Section 2 we revisit the theory of scalar cosmological perturbations for the Universe filled with the
above-mentioned ingredients. We reduce the system of Einstein equations for the first-order metric corrections with respect to the homogeneous background to the
only one basic equation. An illustrative example concerning the structure growth is also given here. Then, in Section 3, we remove the reported earlier sham
limitations on the parameter $\omega$ and briefly summarize the main results.

\

\section{Cosmological perturbation theory revisited}

Let us start with the FLRW metric
\be{1} ds^2=a^2\left(d\eta^2-\delta_{\alpha\beta}dx^{\alpha}dx^{\beta}\right),\quad \alpha,\beta=1,2,3\, , \ee
where $a(\eta)$ is the scale factor; $\eta$ is the conformal time; the comoving coordinates are denoted by $x^{\alpha}$, $\alpha=1,2,3$, and it is supposed for
simplicity that the spatial curvature is absent. The corresponding Friedmann equations in the case of the $\Lambda$CDM model supplemented with an additional
perfect fluid characterized by a constant parameter $\omega$ in the linear EoS read:
\ba{2} \frac{3{\mathcal H}^2}{a^2}&=&\kappa\left(\overline{\varepsilon}_{M} + \overline{\varepsilon}_{R} + \overline{\varepsilon}_{X}\right) +
\Lambda\equiv\kappa\sum\limits_I\overline{\varepsilon}_{I},\quad I=M,R,X,\Lambda\, , \ea
and
\ba{3} \frac{2{\mathcal H}'+{\mathcal H}^2}{a^2}=-\kappa\overline{p}_{R} - \kappa\overline{p}_{X} + \Lambda=-\kappa\left(\frac{1}{3}\overline{\varepsilon}_{R} +
\omega\overline{\varepsilon}_{X}\right) + \Lambda=-\kappa\sum\limits_I\omega_I\overline{\varepsilon}_{I}\, , \ea
where ${\mathcal H}\equiv a'/a\equiv (da/d\eta)/a$; the prime denotes the derivative with respect to $\eta$; $\kappa\equiv 8\pi G_N/c^4$ ($c$ is the speed of
light and $G_N$ is the Newtonian gravitational constant); $\varepsilon_M$, $\varepsilon_R$ and $\varepsilon_X$ represent the energy densities of the
nonrelativistic pressureless matter, radiation and above-mentioned additional component, respectively. The corresponding pressures $p_M$, $p_R$ and $p_X$ satisfy
the following linear equations of state:
\be{4} p_{M}=0,\quad p_{R}=\frac{1}{3}\varepsilon_{R},\quad p_{X}=\omega\varepsilon_{X}\quad\Leftrightarrow\quad p_I=\omega_I\varepsilon_I\, , \ee
where $\omega_M=0$, $\omega_R=1/3$ and $\omega_X\equiv\omega$. Further, the overline indicates the average value, and $\Lambda$ is the cosmological constant (the
corresponding energy density and pressure read: $\varepsilon_{\Lambda}=\overline\varepsilon_{\Lambda}\equiv\Lambda/\kappa$ and $p_{\Lambda}=\overline
p_{\Lambda}=-\Lambda/\kappa$, so $\omega_{\Lambda}=-1$). Throughout the paper the extended notation in formulas is combined with the contracted one, which
contains the subscript $I$ and, generally speaking, is valid for the Universe filled with an arbitrary number of perfect fluids with constant parameters in linear
equations of state. In other words, the contracted notation is valid for the pressureless matter, radiation and an arbitrary number of additional $X$-components
of the specified form.

It should be noted that we do not strive for replacing the $\Lambda$-term by the $X$-component, for example, in order to assure the late-time acceleration of the
Universe expansion. On the contrary, the full range of values of the parameter $\omega$ is studied, including those which do not give rise to the acceleration by
themselves (so in this case the nonzero $\Lambda$-term is still required in order to be in agreement with the observations). At the same time we do not exclude a
possibility $\omega=-1$ (in this case there is no need to introduce the $\Lambda$-term separately, so one should suppose that $\Lambda\equiv0$).

Following the analysis of scalar cosmological perturbations in \cite{Bardeen,Ruth,Mukhanov,Rubakov}, let us consider the metric
\be{5} ds^2\approx a^2\left[\left(1 + 2\Phi\right)d\eta^2 - \left(1-
2\Phi\right)\delta_{\alpha\beta} dx^{\alpha}dx^{\beta}\right]\, ,
\ee
where the so-called conformal-Newtonian gauge is chosen. This particular choice is characterized by the coincidence of the introduced scalar perturbation $\Phi$
as well as the energy-momentum fluctuations $\delta T_i^k$ (see below) with the corresponding gauge-invariant quantities. Then the Einstein equations for the
function $\Phi$ read:
\be{6} \triangle\Phi-3{\mathcal H}(\Phi'+{\mathcal H}\Phi) = \frac{1}{2}\kappa
a^2{\delta T}_{0}^0\, , \ee
\be{7} \frac{\partial}{\partial x^{\alpha}}(\Phi'+{\mathcal H}\Phi)=\frac{1}{2}\kappa a^2{\delta T}_{\alpha}^0\, , \ee
\ba{8} \left[\Phi''+3{\mathcal H}\Phi'+\left(2{\mathcal H}'+{\mathcal
H}^2\right)\Phi\right]\delta_{\alpha\beta}=-\frac{1}{2}\kappa a^2{\delta T}_{\beta}^{\alpha}\, . \ea

The average mixed components $T_i^k$, $i,k=0,1,2,3$, of the total energy-momentum tensor for the investigated multicomponent perfect fluid (pressureless
matter + radiation + X + $\Lambda$) read:
\be{9} \overline{T}_{0}^{0} = \overline{\varepsilon}_{M} + \overline{\varepsilon}_{R} + \overline{\varepsilon}_{X} +
\overline{\varepsilon}_{\Lambda}\equiv\sum\limits_I\overline\varepsilon_I,\quad \overline{T}_{\alpha}^{0}=0\, ,\ee
\ba{10} \overline{T}_{\beta}^{\alpha} &=& -\left(\overline{p}_{R}+\overline{p}_{X}+\overline{p}_{\Lambda}\right)\delta_{\alpha\beta} \nn\\
&=& -\left(\frac{1}{3}\overline{\varepsilon}_{R} + \omega\overline{\varepsilon}_{X}-\overline{\varepsilon}_{\Lambda}\right)\delta_{\alpha\beta}=
-\delta_{\alpha\beta}\sum\limits_I\omega_I\overline{\varepsilon}_{I}, \ea
while for the corresponding fluctuations we have
\ba{11} \delta T_{0}^{0} &=& \delta\varepsilon_{M} + \delta\varepsilon_{R} + \delta\varepsilon_{X}\equiv\sum\limits_I\delta\varepsilon_{I},\quad \nn\\
\delta T_{\alpha}^{0} &=& -\sum_{I}(1+\omega_{I})\frac{\partial \zeta_{I}}{\partial x^{\alpha}}\, ,\ea
\ba{12} \delta T_{\beta}^{\alpha} = -\left(\delta p_{R} + \delta p_{X}\right)\delta_{\alpha\beta} = -\left(\frac{1}{3}\delta{\varepsilon}_{R} +
\omega\delta{\varepsilon}_{X}\right)\delta_{\alpha\beta}=-\delta_{\alpha\beta}\sum\limits_I\omega_I\delta\varepsilon_I\, . \ea
Here, obviously, $\delta\varepsilon_{\Lambda}\equiv0$, so $\delta p_{\Lambda}=-\delta\varepsilon_{\Lambda}=0$. Of course, these equalities by themselves do not
exclude the case of an inhomogeneous perfect fluid with the vacuum-like EoS: its role can be played by an additional $X$-component with the appropriate parameter
$\omega=-1$ and nonzero fluctuations $\delta\varepsilon_{X}\neq0$ and $\delta p_X=-\delta\varepsilon_{X}\neq0$. As regards the introduced quantities $\zeta_I$,
which are treated as importing the first order of smallness, $\nabla\zeta_I$ stands for the gradient part of the spatial vector $\varepsilon_I{\bf v}_I$, where
${\bf v}_I$ is the comoving velocity field of the corresponding $I$-th constituent of the Universe.

In addition, in concordance with \cite{Eingorn} (see also Refs. therein along with \cite{Baumann}), we reject the generally accepted assumption of the linear
relativistic perturbation theory that the energy density fluctuations $\delta\varepsilon_I$ are much less than the corresponding average values
$\overline\varepsilon_I$. In other words, we do not require fulfilment of the inequalities $|\delta\varepsilon_I|\ll\overline\varepsilon_I$, allowing the
fluctuations $\delta\varepsilon_I$ to be nonlinear, in contrast to the textbook material. This is an indispensable step in the direction of elaborating a
relativistic formalism, which would incorporate nonlinear effects at small distances (where $\varepsilon_I$ can essentially exceed $\overline\varepsilon_I$),
while being valid at large distances as well. Meanwhile, following \cite{Eingorn}, we take advantage of the weak gravitational field limit and keep only
first-order (linear) deviations of the metric coefficients from their background values in the Einstein equations. The corresponding retained terms (like, e.g.,
$\triangle\Phi$) clearly predominate over disregarded second-order (nonlinear) terms (like $\Phi\triangle\Phi$) at all scales since $|\Phi|\ll1$.

It is common knowledge that scalar, vector and tensor perturbations are uncoupled at the level of first-order metric corrections. Hence, they can be analyzed
separately. The current paper is entirely devoted to the scalar sector (see \cite{Kiefer} for inclusion of vector modes). In this connection, we consider only the
gradient part of $\delta T_{\alpha}^{0}$ in \rf{11}. The omitted vector part of $\delta T_{\alpha}^{0}$ plays the role of a source for vector perturbations
\cite{Kiefer}, which are beyond the scope of our current investigation. It is worth mentioning that the neglect of the first-order vector perturbations at
arbitrary spatial scales is a common practice \cite{Baumann,Clarkson1}. At large enough distances (in the region of linear energy density fluctuations) the vector
modes decay \cite{Ruth,Mukhanov,Rubakov}, while at sufficiently small distances (in the region of nonlinear clustering) their contributions are subdominant (for
instance, the vector perturbation produced by discrete particles belongs to the post-Newtonian approximation \cite{Eingorn}).

Substituting \rf{11} and \rf{12} into \rf{6}--\rf{8}, we obtain
\ba{13} \triangle\Phi-3{\mathcal H}(\Phi'+{\mathcal H}\Phi) = \frac{1}{2}\kappa a^2\left(\delta\varepsilon_{M} + \delta\varepsilon_{R} + \delta\varepsilon_{X}
\right)=\frac{1}{2}\kappa a^2\sum\limits_I\delta\varepsilon_{I}\, , \ea
\ba{14} \Phi'+{\mathcal H}\Phi=-\frac{1}{2}\kappa a^2\sum_{I}(1+\omega_{I})\zeta_{I}\, , \ea
\ba{15} \Phi'' + 3{\mathcal H}\Phi'+\left(2{\mathcal H}'+{\mathcal H}^2\right)\Phi &=& \frac{1}{2}\kappa a^2\left(\frac{1}{3}\delta{\varepsilon}_{R} +
\omega\delta{\varepsilon}_{X}\right) \nn\\
&=& \frac{1}{2}\kappa a^2\sum\limits_I\omega_I\delta{\varepsilon}_{I}\, . \ea

Substitution of \rf{14} into \rf{13} gives
\ba{16} \triangle\Phi =\frac{1}{2}\kappa a^2\sum\limits_I\delta\varepsilon_{I}-\frac{3}{2}\kappa a^2{\mathcal H}\sum_{I} (1+\omega_{I})\zeta_{I}\, . \ea

In order to find $\Phi$ as a solution of Eq.~\rf{16}, we need, in particular, to determine the quantities $\delta\varepsilon_I$, $I=M,R,X$. For this purpose, let
us analyze the well-known background energy conservation equations
\be{17} \overline\varepsilon_{M}'+3\frac{a'}{a}\overline\varepsilon_{M} = 0 \quad\Rightarrow\quad \overline\varepsilon_{M}\sim\frac{1}{a^3}\, ,
\ee
\be{18} \overline\varepsilon_{R}' + 3\frac{a'}{a}\left(\overline\varepsilon_{R} + \overline p_{R}\right) = \overline\varepsilon_{R}' +
4\frac{a'}{a}\overline\varepsilon_{R} = 0 \quad\Rightarrow\quad \overline\varepsilon_{R}\sim\frac{1}{a^4}\, , \ee
\ba{19} \overline\varepsilon_{X}'+3\frac{a'}{a}\left(\overline\varepsilon_{X} + \overline p_{X}\right) = \overline\varepsilon_{X}' + 3(1 +
\omega)\frac{a'}{a}\overline\varepsilon_{X} = 0 \quad \Rightarrow \quad \overline\varepsilon_{X}\sim\frac{1}{a^{3(1+\omega)}}\, , \ea
as well as the perturbed energy-momentum conservation equations (see, e.g., \cite{Rubakov} for their linearized form):
\ba{20} \delta\varepsilon_I' + 3\frac{a'}{a}\left(\delta\varepsilon_I+\delta p_I\right) -
3\left(\overline{\varepsilon}_I+\overline{p}_I\right)\Phi'+\nabla\left[\left({\varepsilon}_I+{p}_I\right){\bf v}_I\right]=0\, , \ea
where $\nabla\left[\left({\varepsilon}_I+{p}_I\right){\bf v}_I\right]=\left(1+\omega_I\right)\triangle\zeta_I$, and
\be{21} \left(1+\omega_I\right)\zeta'_I+ 4\frac{a'}{a}\left(1+\omega_I\right)\zeta_I+ \delta p_I+\left(\overline{\varepsilon}_I+\overline{p}_I\right)\Phi=0\, .\ee
Really, we study solely noninteracting pressureless matter, radiation and the $X$-component; therefore, Eqs.~\rf{20} and \rf{21} are valid for each constituent of
the Universe separately. In addition, we have dropped the contributions containing both $\delta\varepsilon_I$ and $\Phi$ since these terms would import the second
order of smallness in the Einstein equations and therefore lie beyond the accuracy adopted here. In other words, we strive for finding such expressions for
$\delta\varepsilon_I$, $I=M,R,X$, which would conform with the precision inherent in Eqs. (13)--(15) and have no second-order trace (see \cite{Eingorn,BrilEin}
for additional confirmative reasoning regarding the $M$-component). From \rf{20} we get
\ba{22} &{}& \delta\varepsilon_{M}' + 3\frac{a'}{a}\delta\varepsilon_{M}  - 3\overline\varepsilon_{M}\Phi'+\triangle \zeta_M= 0
\quad\Rightarrow\quad\nn\\
&{}& \delta\varepsilon_{M} = \frac{\delta A_M}{a^{3}} + 3\overline\varepsilon_{M}\Phi,\quad \delta A'_M=-a^3\triangle \zeta_M\, , \ea
\ba{23} &{}& \delta\varepsilon_{R}' + 4\frac{a'}{a}\delta\varepsilon_{R}  - 4\overline\varepsilon_{R}\Phi' +\frac{4}{3}\triangle \zeta_R= 0
\quad\Rightarrow\quad\nn\\
&{}&  \delta\varepsilon_{R} = \frac{\delta A_{R}}{a^{4}} + 4\overline\varepsilon_{R}\Phi,\quad \delta A'_R= -\frac{4}{3}a^4\triangle \zeta_R\, ,\ea
\ba{24} &{}& \delta\varepsilon_{X}' + 3(1+\omega)\frac{a'}{a}\delta\varepsilon_{X}  - 3(1+\omega)\overline\varepsilon_{X}\Phi'
+(1+\omega)\triangle \zeta_X= 0 \quad\Rightarrow\quad\nn\\
&{}& \delta\varepsilon_{X} = \frac{\delta A_{X}}{a^{3\left(1+\omega\right)}} + 3\left(1+\omega\right)\overline\varepsilon_{X}\Phi,\quad \delta
A'_X=-(1+\omega)a^{3(1+\omega)}\triangle \zeta_X\, .\ea

Thus, the sought-for fluctuations $\delta\varepsilon_I$, $I=M,R,X$, are determined. These energy density perturbations, along with the metric scalar perturbation
$\Phi$, related to the chosen conformal-Newtonian gauge, coincide with the corresponding gauge-independent quantities (see, e.g., \cite{Mukhanov}). Hence, the
same applies to the introduced functions $\delta A_I$, which actually represent linear combinations of $\delta\varepsilon_I$ and $\Phi$. In confirmation of the
result \rf{22} for $\delta\varepsilon_{M}$ let us mention the important fact that this result exactly coincides with what follows directly from the well-known
formula $\varepsilon_M=\rho_{M}c^2\left[g_{00}/(-g)\right]^{1/2}$ \cite{Landau} ($\overline\varepsilon_M=\overline\rho_{M}c^2/a^3$ and $\delta A_M=\delta\rho_M
c^2$, where $\rho_M$ and $\overline\rho_M$ represent the rest mass density of pressureless matter in comoving coordinates and its average value, respectively,
while $\delta\rho_M\equiv\rho_M-\overline\rho_M$). The derived expressions \rf{22}--\rf{24} can be certainly united:
\ba{25} \delta\varepsilon_{I} = \frac{\delta A_{I}}{a^{3\left(1+\omega_I\right)}} + 3\left(1+\omega_I\right)\overline\varepsilon_{I}\Phi,\quad  \delta
A'_I=-(1+\omega_I)a^{3(1+\omega_I)}\triangle \zeta_I\, ,\ea
and this general expression is also valid for $I=\Lambda$ since $\delta A_{\Lambda}\equiv0$. Substituting \rf{25} into Eq.~\rf{16}, we get
\ba{26} \triangle\Phi - \frac{3}{2}\kappa a^2\left[\sum\limits_I\left(1+\omega_I\right)\overline\varepsilon_I\right]\Phi &=& \frac{1}{2}\kappa
a^2\sum\limits_I\frac{\delta A_{I}}{a^{3\left(1+\omega_I\right)}} \nn\\
&-&\frac{3}{2}\kappa a^2{\mathcal H}\sum_{I}(1+\omega_{I}) \zeta_{I}\, . \ea

Thus, we arrive at determination of the scalar perturbation $\Phi$ by the fluctuations $\delta A_I$, $I=M,R,X$, describing the ``intrinsic'' perturbations of the
corresponding energy densities $\varepsilon_I$ (the first terms in \rf{22}--\rf{24}, respectively), and the quantities $\zeta_I$. The second term on the left-hand
side of Eq.~\rf{26} is directly proportional to the sum of ``responses'' of the energy densities $\varepsilon_I$ to the presence of the inhomogeneous
gravitational field (the second terms in \rf{22}--\rf{24}). The expression
\be{27} \frac{3}{2}\kappa \left[\sum\limits_I\left(1+\omega_I\right)\overline\varepsilon_I\right]= \frac{3\left(\mathcal{H}^2-\mathcal H'\right)}
{a^2}\equiv\frac{1}{\lambda^2} \ee
determines the Yukawa interaction range $\lambda$ (in physical coordinates), which is universal for each component, in complete agreement with the corresponding
statement made in \cite{Eingorn}. Really, the solution of Eq.~\rf{26} has the form
\be{28} \Phi=\Phi_M+\Phi_R+\Phi_X\equiv\sum\limits_I\Phi_I,\quad \Phi_{\Lambda}\equiv0\, ,\ee
where each contribution $\Phi_I$ satisfies the equation
\ba{29} \triangle\Phi_I - \frac{a^2}{\lambda^2}\Phi_I= \frac{\kappa}{2a^{1+3\omega_I}}\delta A_{I}-\frac{3}{2}\kappa a^2{\mathcal H} (1+\omega_{I})\zeta_{I}\, .
\ea
Finding $\Phi_I$ from \rf{29}, one can substitute the result into \rf{28} and, hence, find the total scalar perturbation $\Phi$, provided that the sources $\delta
A_I$ and $\zeta_I$ are known. It is interesting that the formulated Yukawa range definition \rf{27} holds true in the case of varying parameters $\omega_I$ as
well (or, in other words, in the case of nonlinear equations of state) if linear perturbations are at the center of attention. Indeed, this follows directly from
the fact that the ``response'' $\delta\varepsilon_I=3\left(\overline{\varepsilon}_I+\overline{p}_I\right)\Phi$ still satisfies the equation $\delta\varepsilon_I'
+ 3\mathcal{H}\left(\delta\varepsilon_I+\delta p_I\right) - 3\left(\overline{\varepsilon}_I+\overline{p}_I\right)\Phi'=0$ for the EoS $p_I=f(\varepsilon_I)$ with
an arbitrary nonlinear function $f$, as one can easily prove with the help of the evident equalities $\overline p_I=f(\overline\varepsilon_I)$, $\delta
p_I=(d\overline p_I/d\overline\varepsilon_I)\delta\varepsilon_I$, and $\overline\varepsilon'_I+3\mathcal{H}(\overline{\varepsilon}_I+\overline{p}_I)=0$. The
perfect fluid with Chevallier-Polarski-Linder parametrization of the EoS ($\omega$ as a linear function of $a$) and Chaplygin gas represent concrete popular
examples, which belong to the discussed class of supposed Universe components.

It is not difficult to show that the remaining Einstein equations \rf{14} and \rf{15} are satisfied. For example, one can act by the Laplace operator $\triangle$
on both sides of Eq.~\rf{14} and express $\triangle\Phi$ from \rf{26}, $\delta A'_I$ from \rf{25} and $\zeta'_I$ from \rf{21} as well as use the Friedmann
equations \rf{2} and \rf{3}. Then, in order to prove Eq.~\rf{15}, one can start with expressing $\Phi'$ from \rf{14}.

Therefore, the initial system of Eqs.~\rf{13}--\rf{15} has been reduced to the only one equation \rf{26} supplemented with perturbed energy-momentum conservation
equations. Owing to \rf{26}, the spatial averaging of $\Phi$ gives the zero value $\overline{\Phi}=0$ (which is highly desirable \cite{EBV}), so there are no
first-order backreaction effects, as expected from the very beginning (for demonstration technique see \cite{Eingorn}). It is noteworthy as well that, as we
emphasized above, fulfilment of the inequalities $|\delta\varepsilon_I|\ll\overline\varepsilon_I$ is not demanded.

Eq.~\rf{26} for the scalar perturbation $\Phi$ represents our main novel result. Its derivation became possible due to splitting of $\delta\varepsilon_I$ into
``intrinsic'' fluctuations and ``responses'' to the presence of the inhomogeneous gravitational field (see the first and second terms in \rf{25}, respectively).
First, this splitting helps to find $\Phi$ in the whole space for linear as well as nonlinear perturbations of energy densities, without $1/c$ series expansion,
``dictionaries'' or admixtures of second-order quantities (in contrast to, e.g., \cite{Adamek}). In other words, we have not made any extra assumptions
supplementing the weak field limit. Second, separation of intrinsic fluctuations reveals the Yukawa nature of their gravitation, corroborating the ideas reported
earlier in \cite{Eingorn} (see also \cite{Eingorn2017}) for a much wider class of cosmological scenarios.

Returning momentarily to the linear relativistic perturbation theory, it is very interesting to investigate the structure growth in terms of intrinsic
fluctuations. For the illustration purposes, let us restrict ourselves to the matter-dominated stage of the Universe evolution, disregarding all other
constituents. Then, being based on \rf{21}, \rf{22} and \rf{29}, it is not difficult to demonstrate that the Fourier transform $\hat{\delta\rho}_M(\eta,{\bf k})$
of the rest mass density perturbation $\delta\rho_M(\eta,{\bf r})$ satisfies the following equation:
\ba{31} \hat{\delta \rho}''_M + \mathcal{H}\cfrac{1+2\chi}{1+\chi}\hat{\delta \rho}'_M- \cfrac{1}{1+\chi}\frac{\kappa\overline\rho_{M}c^2}{2a}\hat{\delta\rho}_{M}
= 0\, ,\ea
where $\chi\equiv a^2/\left(k^2\lambda^2\right)$. Now, if $a/k\ll\lambda$ (small scales as compared with the Yukawa interaction range), $\chi\ll1$, and then
\ba{32} \hat{\delta \rho}''_M + \mathcal{H}\hat{\delta \rho}'_M- \frac{\kappa\overline\rho_{M}c^2}{2a}\hat{\delta\rho}_{M} = 0\, ,\ea
admitting the growing mode $\hat{\delta\rho}_M\sim a$, as it is generally known. One should remember that the rest mass densities $\rho_M$ and
$\rho^{\mathrm{ph}}_M$ in comoving and physical coordinates, respectively, are interconnected by means of the relationship $\rho^{\mathrm{ph}}_M=\rho_M/a^3$. The
average physical rest mass density $\overline\rho^{\mathrm{ph}}_M$ behaves as $1/a^3$, while $\overline\rho_M=\mathrm{const}$. The density contrast
$\delta_M\equiv \delta\rho^{\mathrm{ph}}_M / \overline\rho^{\mathrm{ph}}_M\equiv \delta\rho_M / \overline\rho_M\sim\delta\rho_M$.

However, in the opposite case $a/k\gg\lambda$ (large scales as compared with the Yukawa range), $\chi\gg1$, and therefore
\ba{31} \hat{\delta \rho}''_M + 2\mathcal{H}\hat{\delta \rho}'_M= 0\, ,\ea
admitting the constant mode $\hat{\delta\rho}_M=\mathrm{const}$. The issued statement remains unchanged if the cosmological constant $\Lambda$ is also taken into
consideration along with the nonrelativistic matter. This is one more novel result obtained in the current paper, which can be formulated as follows: the
evolution of intrinsic fluctuations (the rest mass density perturbation in our example) is suppressed at distances exceeding the Yukawa range of gravitational
interaction. Of course, this assertion is expected right after the introduction of this range. Really, from the physical point of view, inhomogeneities, which are
separated by the distance greater than $\lambda$, gravitationally almost do not ``feel'' each other. Consequently, it is hard for them to participate in the
formation of the same structure.

One should emphasize that the revealed upper bound for a spatial domain of the probable structure growth corroborates the hypothesis formulated in \cite{Eingorn}:
the established Yukawa range bears a direct relation to the explanation of existence of the largest known cosmic structures \cite{wall,wall2,ring,LQG} (according
to \cite{Eingorn}, $\lambda\approx3.7$ Gpc at present, and their dimensions do not exceed this limiting value).

\

\section{Conclusion}

The following main results, obtained in the present paper, deserve mentioning:

--- the first-order scalar perturbation $\Phi$ is determined at arbitrary (sub-horizon and super-horizon)
scales by means of the Helmholtz equation \rf{26} in the weak gravitational field limit;

--- the velocity-independent part of $\Phi$ is characterized by the finite time-dependent Yukawa interaction range $\lambda$, defined by the formula \rf{27} and
being the same for each supposed component of the inhomogeneous Universe;

--- Yukawa nature of gravitational interaction between inhomogeneities leads to suppression of the structure formation and evolution at distances greater
than $\lambda$.

Let us also enumerate two other important corollaries from the developed description of scalar cosmological perturbations. First, in the framework of the
$\Lambda$CDM model supplemented with an additional perfect fluid characterized by a constant parameter $\omega$ in the linear EoS (or, generally speaking, with an
arbitrary number of such fluids) there is no theoretical limitation imposed on $\omega$ at the level of Einstein equations, as distinct from what was reported
before in \cite{BEZ1}. In that paper only two values $\omega=-1$ (the vacuum-like EoS) and $\omega=-1/3$ survived. Now these severe restrictions are removed
irrevocably. Second, the energy-momentum conservation takes place for each Universe constituent separately, and there is no need at all in any additional
radiation contribution exchanging momentum with nonrelativistic matter, which was artificially introduced in \cite{EZcosm2}.

Summarizing, we have made the promised direct generalization of the approach elaborated in \cite{Eingorn} for the concordance cosmological model and discrete
presentation of matter sources to the case of nonconventional models with extra perfect fluids and continuous (hydrodynamical) presentation. This broad extension
of the preceding achievements advocates linear equations for the first-order metric corrections (this is absolutely logical for weak fields). Evidently, these
equations are numerically solvable with lesser effort and higher accuracy than sophisticated equations containing selected nonlinear admixtures, which are implied
by the dictionary-based approach (see \cite{Adamek} and Refs. therein). Being based on our results, it is quite possible to construct similarly an appropriate
second-order scheme for arbitrary scales (see \cite{BrilEin} for the case of the standard $\Lambda$CDM components without supplementary fluids) and investigate
the backreaction effects. This feasible scheme would be at an advantage over the second-order extension of the linear relativistic perturbation theory (see, e.g.,
\cite{observ1} for the application of the latter to observable quantities), since all relevant small-scale effects would be properly taken into account.
Corresponding hydrodynamical simulations of the structure growth and related investigations can enable us to distinguish among the $\Lambda$CDM paradigm and its
various competing alternatives.

\

\section*{Acknowledgements}

The work of M.~Eingorn was partially supported by NSF CREST award HRD-1345219 and NASA grant NNX09AV07A.

The work of R.~Brilenkov was partially supported by the EMJMD Student Scholarship from the Erasmus\,+\,: Erasmus Mundus Joint Master Degree programme AstroMundus
in Astrophysics.

\


\end{document}